\documentstyle[prb,aps,twocolumn]{revtex}
\begin{document}
\wideabs{
\title{Valence band excitations in V$_2$O$_5$}
\author{S. Atzkern, S.V. Borisenko, M. Knupfer, M.S. Golden, J. Fink}
\address{Institut f\"ur Festk\"orper- und Werkstofforschung Dresden, P.O.Box 270016, D-01171 Dresden, Germany}
\author{A.N. Yaresko \cite{imp}, V.N. Antonov \cite{imp}}
\address{Max-Planck-Institut f\"ur Physik komplexer Systeme, N\"othnitzer Str. 38, D-01187 Dresden, Germany}
\author{M. Klemm, S. Horn}
\address{Institut f\"ur Physik, Universit\"at Augsburg, D-86159 Augsburg, Germany}
\date{\today}
\maketitle
\begin{abstract}

We present a joint theoretical and experimental investigation of the electronic and optical properties of vanadium pentoxide. Electron energy-loss spectroscopy in transmission was employed to measure the momentum-dependent loss function. This in turn was used to derive the optical conductivity, which is compared to the results of band
structure calculations.
A good qualitative and quantitative agreement between the theoretical and the experimental optical conductivity was observed.
The experimentally observed anisotropy of the optical properties of V$_2$O$_5$ could be understood in the light of an analysis of the theoretical data involving the decomposition of the calculated optical conductivity into contributions from transitions into selected energy regions of the conduction band. In addition, based upon a tight binding fit to the band structure, values are given for the effective V~3$d_{xy}$--O~2$p$ hopping terms and are compared to the corresponding values for $\alpha '$--NaV$_2$O$_5$.
\\
\\
\end{abstract}

}
\section{Introduction}
\label{introd}

The transition metal element vanadium with the electronic configuration [Ar]3d$^3$4s$^2$ builds a variety of binary and ternary compounds. 
As vanadium can be found in the same compound in two different oxidation states, stoichiometries with many different non-integer valencies are possible. In the Magn\'eli phases V$_n$O$_{2n-1}$ (3 $\leq n \leq$ 8), for example, different combinations of V$^{3+}$ and V$^{4+}$ provide us with the formal valencies $4-\frac{2}{n}$, whereas in the compounds V$_n$O$_{2n+1}$ (2 $\leq n \leq$ 4, n = 6), the formal valencies $4+\frac{2}{n}$ are realized.
Almost all binary vanadates exhibit a metal-insulator transition (MIT), but in most cases the discussion about the driving force for the MIT is controversial. Even the origin of the MIT in V$_2$O$_3$ which is one of the best studied Mott-Hubbard systems was recently called into question. \cite{Ezhov}

In the binary vanadate V$_2$O$_5$ with the formal electronic $d^0$ configuration there is no evidence for a MIT.\cite{sipr} However, its applications as e.g. a gas sensor, a catalyst, or the cathode material in rechargeable lithium batteries makes this compound interesting (see for instance Refs. \onlinecite{micocci,haber,yamaki}). 
Recently V$_2$O$_5$ has attracted attention due to its electronic and structural relationship to the ternary $\alpha '$-NaV$_2$O$_5$ in which the vanadium atoms are formally mixed-valent (V$^{4.5+}$). \cite{goering,Eyert} The dynamics of the V~3$d$-electrons in $\alpha '$-NaV$_2$O$_5$ are the subject of controversial discussion and hence the investigation of V$_2$O$_5$ might supply us with a good basis for a deeper understanding of the electronic properties of $\alpha '$-NaV$_2$O$_5$.

In the present work the optical properties of V$_2$O$_5$ single crystals have been investigated using electron energy-loss spectroscopy (EELS) in transmission and first-principles local-density approximation calculations. We found that the energy and momentum dependent loss spectra show dispersionless features, which reflects the local character of the  transitions involved. Considering the crystal structure of vanadium pentoxide as a three-dimensional network of corner- and edge-sharing distorted VO$_6$ octahedra, we described the basic properties of the electronic structure in terms of bonding-antibonding and ligand-field splittings. The calculated optical conductivity exhibits a strong anisotropy which is in an excellent agreement with our experimental results. We have identified the possible origin of the main dipole transitions by selecting three energy intervals in the conduction band, corresponding to states with pronounced orbital symmetry and by investigating the contribution of the onsite O~2$p~\rightarrow$~O~3$d$ transitions to the optical conductivity. 

\section{Methodology}
\label{method}

\subsection{Samples}
\label{samp}

Single crystals of V$_2$O$_5$ were grown from the melt. The detailed procedure is described elsewhere.\cite{goering} The bright yellow crystals were up to \mbox{30 mm} in diameter and \mbox{0.2 mm} thick.
V$_2$O$_5$ has an orthorhombic structure belonging to the space group P$mmn$ with the lattice constants \mbox{a = 11.51 \AA}, \mbox{b = 3.56 \AA} and \mbox{c = 4.37 \AA}.\cite{enjalbert} In Fig. \ref{Structure} the crystal structure is depicted in three different ways. In the upper panel the vanadium atoms (filled black spheres) and the three inequivalent oxygen positions O$_b$ (bridge), O$_c$(chain) and O$_v$(vanadyl) are shown together with the bonds between the vanadium and the nearest neighbor oxygen atoms. In the $\bf b$ direction, linear chains are built up from alternating vanadium and chain oxygen atoms. In the $\bf a$ direction, the vanadium atoms are connected via the bridge oxygens. This arrangement can be described as a ladder structure with the legs running along $\bf b$ and the rungs along 
$\bf a$. 

Along the $\bf c$ direction the vanadyl oxygens are located above and below the vanadium atoms creating the shortest (1.576 \AA) and longest (2.793 \AA) vanadium-oxygen distances in this structure.
The other V-O distances are 1.776 \AA\ (V-O$_b$), 1.878 \AA\ (V-O$_c$) along the leg and 2.018 \AA\ (V-O$_c$) between legs of neighboring ladders. Each vanadium atom and its five nearest oxygen neighbors create VO$_5$-pyramids which share their corners within the ladder and their edges between neighboring ladders (see lower right panel of Fig. 1). The resulting layers are stacked along the $\bf c$ direction. We would like to remark that actually the vanadium is shifted out of the base plane of the pyramid toward the vanadyl oxygen. If we complete the pyramids to form strongly distorted octahedra by adding the vanadyl oxygen from the lower lying pyramid (2.793 \AA\ from the V atom) we obtain the third alternative description of the crystal structure in terms of corner and edge-sharing VO$_6$-octahedra (lower left panel of Fig. \ref{Structure}).

The layered pyramid structure is ideal to explain the good cleavage behaviour in the ab-plane. The crystal cleavage results in dangling bonds at the oxygen and vanadium sites which are the origin of the properties which makes V$_2$O$_5$ interesting in applications as a catalyst or gas sensor.\cite{haber} However, for the description of the electronic structure of vanadium pentoxide the structure model with the VO$_6$ octahedra and the resulting ligand field splitting of the V~3$d$ states will be more useful.

For the measurements using electron energy-loss spectroscopy in transmission, thin films of about 1000 \AA\ thickness were cut from the single crystals with a diamond knife using an ultramicrotome. Because of the good cleavage behavior, the crystallinity remains conserved after cutting parallel to the ab-plane, which is in contrast to the behavior for any cutting plane containing the $\bf c$ direction. The high quality and orientation of the single crystalline samples were checked by {\it in~situ} electron diffraction.

\subsection{Experiment}
\label{exper}
EELS in transmission with a primary beam energy of 170 keV was performed on free standing films at room temperature (for experimental details see Ref.\onlinecite{fink}). The energy and momentum transfer (q) resolution were chosen to be 110 meV and 0.5 \AA $^{-1}$ for q $\leq$ 0.4 \AA $^{-1}$, and 160 meV and 0.6 \AA $^{-1}$ for q $>$ 0.4 \AA $^{-1}$ due to the decrease of the cross section at higher momentum transfer. 

EELS in transmission provides us with the momentum and energy dependent loss function Im(-1/$\varepsilon$($\bf q$, $\omega$)), from which, by means of the Kramers-Kronig relations, the real part of the negative inverse dielectric function and thus all the optical constants such as for example the optical conductivity $\sigma ({\bf q},\omega)$ can be calculated. For small momentum transfer only dipole transitions are allowed and for the limit q = 0 the transition matrix elements are the same as in optics. For the Kramers-Kronig analysis the loss spectra closest to the optical limit (q = 0.1 \AA$^{-1}$) were used in order to derive the optical conductivity spectra. 

\subsection{Computational details}
\label{calcul}

The electronic structure of V$_2$O$_5$ was calculated self-consistently within the framework of the local-density approximation (LDA) to the density-functional theory \cite {Hohenberg,Kohn}, using the linear muffin-tin orbital (LMTO) method \cite{Andersen} with the so-called combined correction term \cite{Andersen2} taken into account. For the exchange-correlation potential the von Barth-Hedin parametrization \cite{bh72} has been employed. The
calculations were semi-relativistic, i.e. all relativistic effects except for the spin-orbit coupling were taken into account. 

Since the crystal structure of V$_2$O$_5$ is loosely packed which leads to a large overlap of the atomic spheres we inserted 24 additional empty spheres into the simple orthorhombic unit cell of V$_2$O$_5$. Their positions and radii were taken from Ref. \onlinecite{Eyert}. The angular momentum expansion of the basis functions included $l=3$ for vanadium and $l=2$ for oxygen and the empty spheres. The O~3$d$ states have only a minor effect on the energy bands. However, they give a significant contribution to the optical conductivity due to the large oscillator strength of the corresponding $p\rightarrow d$ transitions (see Sec. \ref{optics}) . The Brillouin zone (BZ) integrations in the self-consistency loop were performed using the improved tetrahedron method \cite{blochl94} on a grid of 864 $\bf k$ points.

The real part of the optical conductivity tensor components was computed from the energy bands and the LMTO eigenvectors on the basis of the linear-response expressions.\cite{kubo57,maks88,rmme93} The dispersive part
of the conductivity tensor was obtained via the Kramers-Kronig transformation. Finite-lifetime effects and experimental resolution were simulated by broadening the calculated spectra with a Lorentzian of width 0.3~eV.

The calculated band gap is somewhat less than the experimentally observed gap of 2.3~eV. Such an underestimation of the band gap is a well-known limitation of the LDA,\cite{Pickett} which is due to the non-exact treatment of the electron exchange and correlation. Nevertheless this discrepancy causes only a downward shift of the energy scale of the optical spectra. 
As normalization parameters in the Kramers-Kronig analyses of the experimental and theoretical data we used the static dielectric constants $\varepsilon_1(0)$ of 5.57, 4.8 and 4.55 for $\bf a$, $\bf b$ and $\bf c$ directions, respectively, obtained from the zero-frequency limit of the calculated $\varepsilon_1(\omega)$.

\section{Results and discussion}
\label{results}

\subsection{Loss functions}
\label{lof}

In Fig. \ref{Lossfunc}a and \ref{Lossfunc}b the energy dependent loss functions for different momentum transfers parallel to the crystallographic {\bf a} and {\bf b} directions are shown, respectively. The spectra are normalized at higher energy (about 15~eV) where the shape does not change with the momentum transfer. 

At first sight for both directions the spectra look similar. The spectral onset is found to be at \mbox{2.3~eV} which agrees well with optical transmission \mbox{(E$_{gap}$ = 2.24~eV)} \cite{chain}, optical absorption \mbox{(E$_{gap}$ = 2.3~eV)} \cite{cogan} and optical reflectance measurements \mbox{(E$_{gap}$ = 2.38~eV)}.\cite{moshfegh} Two rather narrow peaks at \mbox{3.5~eV} and at \mbox{$\sim$ 9.7~eV} are conspicuous while additional broader features are found to be at \mbox{4.8~eV}, \mbox{$\sim$ 7~eV} and between \mbox{10~eV} and \mbox{12~eV}. The intensity of all these features decreases with increasing momentum transfer whereas an additional feature at 5.1~eV which appears at about \mbox{0.3 \AA$^{-1}$} gains intensity with higher {\bf q}. These results can be explained by the {\bf q}-dependence of the transition matrix elements which means that the former feature can be identified as resulting from interband plasmons related to dipole transitions while the latter feature most likely originates from an interband plasmon related to an optically forbidden quadrupole transition. In both directions all the features show no clear dispersion.
In general a dispersive interband plasmon is observed in EELS when the bands forming both the initial and 
final states have significant dispersion.\cite{fink} In this case, the non-dispersive nature of the EELS features
in V$_2$O$_5$ is attributable to the local, non-dispersive character of the final states, as the  O 2$p$-dominated valence band states do have significant bandwidth.

For a simple interpretation we first consider a purely ionic picture. In the low  energy region the electrons from the completely filled O~2$p$ states are excited into empty V~3$d$ states. In order to estimate qualitatively the character of the lowest unoccupied states we consider the local symmetry around the vanadium atoms. As mentioned in section \ref{samp}, one possible description of the crystal structure is the arrangement of corner and edge-sharing distorted VO$_6$ octahedra (Fig. \ref{Structure}). The octahedral symmetry implies a splitting of the V~3$d$ levels into $t_{2g}$ and energetically higher lying $e_g$ states. Since in the ab-plane the differences between the V-O distances are small whereas the deviations from a regular extension of the octahedra in {\bf c} direction are large we proceed with a tetragonal distortion of the octahedra. The tetragonal ligand field then gives rise to a further lowering of the $d_{xy}$ level and a splitting of the $e_g$ states. Hence the lowest excitations \mbox{(E $\sim$ 3.5~eV)} with momentum transfer in {\bf a} and {\bf b} direction can be related to dipole transitions from the occupied O~2$p$ states into the V~3$d_{xy}$ state.

On very general grounds, one expects that the O 2$p$ states should hybridize with vanadium 3$d$ states to produce a bonding and an energetically higher lying antibonding combination. The bonding states then build the valence band (VB) while the conduction band (CB) is formed by the antibonding states. The magnitude of the splitting between the hybridized states depends on the type of bonding. The $\sigma$ bonding, e.g. between the O$_b$~2$p_x$ and the V~3$d_{x^2-y^2}$ state, causes  a much stronger splitting than the weaker $\pi$ bonding, as occurs between the O$_b$~2$p_y$ and the V~3$d_{xy}$ state.
The difference between transitions from bonding states into $\sigma^*$ states and those into $\pi^*$ states is visible in the EELS spectra as the appearance of the spectral features in two well seperated energy regions. Transitions into the $\pi^*$ states give rise to spectral weight between \mbox{3.5~eV} and \mbox{7~eV}  while the energy losses between \mbox{9~eV} and \mbox{12~eV} correspond to much more intensive transitions into the $\sigma^*$ states.

On closer inspection the loss functions measured with momentum transfer in the $\bf a$ and $\bf b$ direction show small differences in the spectral weight which illustrate the anisotropy of the electronic structure in the ab-plane. This anisotropy is caused by the overlap between different pairs of O2$p$ and V3$d$ orbitals along  the $\bf a$ and $\bf b$ direction, respectively, and will be studied in detail in the next section.

\subsection{Partial densities of states}
\label{subsec:PDOS}
In order to gain a deeper insight into the origin of the interband transitions contributing to the spectra we calculate the partial densities of states for all inequivalent atoms and resolve the partial V~3$d$ and O~2$p$ contributions into their five and three symmetry components, respectively. The upper left panel of Fig. \ref{DOS}(a) shows the total V~3$d$ DOS which is consistent in most features with previously published theoretical studies that used various $\it ab$ $\it initio$ techniques. \cite{Eyert,Chakrab} 
Three groups of bands could be identified in the energy range from -6~eV to 10~eV giving rise to the DOS (the zero point of energy corresponds to the top of the valence band). The valence band (VB) has a width of $\sim$5.3~eV and is separated from the conduction band (CB) by an indirect optical gap of $\sim$1.73~eV between the $\Gamma$ point and a point close to (0, $\pi$ ,$\pi$) in the Brillouin zone of the simple orthorhombic lattice. A narrow conduction sub-band of only 0.47~eV width splits off from the CB resulting in an additional band gap of 0.48~eV. 

In Fig. \ref{DOS} the DOS components projected onto the cubic harmonics ($p_x$, $p_y$, $p_z$, $d_{xy}$, $d_{xz}$, $d_{yz}$, $d_{x^2-y^2}$ and $d_{3z^2-r^2}$) which form the basis of the irreducible representations of the local subgroup of vanadium and oxygen sites are also shown. 
The energy ranges with high partial DOS are marked with a grey background. The present calculations yield an almost pure V~3$d_{xy}$ character for the split-off sub-band.
The V~3$d_{xy}$/3$d_{xz}$/3$d_{yz}$ states appear within the energy regions between -3~eV and 0~eV and between 2~eV and 4~eV whereas the V~3$d_{x^2-y^2}$/3$d_{3z^2-r^2}$ states have dominant contributions in energy regions below -3~eV and above 4~eV (see Fig. \ref{DOS}(a)). For the latter states this is a consequence of the stronger bonding-antibonding splitting resulting from $\sigma$-like V~3$d$ -- O~2$p$ interaction mentioned above. 

The shapes of the O~2$p$ partial DOS and their symmetry components are quite sensitive to the sort of oxygen atoms, thus reflecting the different bonding environments [see Fig. \ref{DOS}(b)]. Keeping in mind that every bridge and vanadyl oxygen atom is shared by two VO$_6$ octahedra we point out the resemblance between the O$_b$~2$p_y$/2$p_z$ and O$_v$~2$p_x$/2$p_y$ partial DOS which are involved in $\pi$ bonding as well as the O$_b$~2$p_x$ and O$_v$~2$p_z$ partial DOS participating in $\sigma$ bonding with vanadium 3$d$ states. 
In contrast, each chain oxygen belongs at once to three distorted octahedra. Such a location suggests that the O$_c$ components create bonds of both types. This fact is reflected in the corresponding DOS which can be considered as the superposition of the two distributions discussed above. 

\subsection{Optical conductivity}
\label{optics}

We have calculated the optical conductivity $\sigma(\omega)$ of V$_2$O$_5$ crystals for energies up to $\sim$27~eV. Since the most important and well pronounced features of the EELS spectra are found to be in the energy range from 0 to 14~eV, we use these limits to present the calculated and experimental results along the two main crystallographic directions, {\bf a} and {\bf b}, in Fig. \ref{Th+exp}. 
In order to achieve the best correspondence, we applied the same upward energy shift of 0.35~eV to both calculated curves, which is most likely due to the underestimation of the band gap (see Sec. \ref{calcul}). The overall agreement between the calculated curves and those derived from the experiment is striking. Along both directions the calculated energy positions of the main spectral features coincide almost precisely with the experimental ones. A strong absorption  at 3~eV  (3.2~eV) for the $\bf a$ ($\bf b$) direction as well as the following drop of the spectral weight  is reproduced by the calculation. The separation of the peak at 4.5~eV from the group of more intense peaks located between 5.5 and 9.0~eV for the $\bf a$ direction and the qualitatively different picture in the given energy interval in the case of the $\bf b$ direction are also reflected in the theoretical curves. The only noticeable discrepancies seem to be an overestimation of the intensity of the peaks at $\sim$7~eV ($\bf a$ direction) and $\sim$5~eV ($\bf b$ direction) and the theoretical underestimation of the energy of the broad features appearing between 10 and 12~eV in the experiment.

The observed correspondence between calculations and experiment allows us to draw additional information from the detailed analysis of the theoretical spectra. For that purpose we present the calculated components of the optical conductivity tensor along the three principal crystallographic axes in Fig. \ref{Sigmas}(a). 

It is difficult to associate the features in the optical conductivity curve with transitions involving specific pairs of bands because of the multiplicity in the manifold of band structures. Therefore we consider contributions to the optical conductivity from transitions between the occupied VB states and selected energy intervals in the unoccupied electronic structure. In Fig. \ref{Sigmas}(a) we display only three of such components since the rest play only a negligible role in the energy region from 0  to 11~eV. The first contribution (area patterned with horizontal lines in Fig. \ref{Sigmas}(a)) corresponds to transitons into empty states located between 0 and 3.06~eV. This interval comprises the bands dominanted by V~3$d_{xy}$ states mixing with O$_b$~2$p_y$ and O$_c$~2$p_x$/2$p_y$. The calculated absorption threshold is about 1.96~eV which is 0.23~eV larger than the minimal indirect gap. The optical absorption increases rapidly immediately above the direct-gap threshold only for the $\bf a$ direction and has pronounced minima at 3.7~eV for the $\bf a$ and $\bf b$ direction. Such a drop in absorption is obviously the consequence of the presence of the additional band gap mentioned above. Our calculations show that the sharp maximum at 2.17~eV in $\sigma_{xx}$ is due to transitions between the bands straddling the chemical potential, i.e. representing the main gap. 

The lowest unoccupied band has already been defined as having almost pure V~3$d_{xy}$ character while the highest occupied band is a mixture of mainly O$_b$~2$p_y$/2$p_z$,  O$_v$~2$p_x$ and O$_c$~2$p_x$ characters (see Fig. \ref{DOS}(b)). 
Since only the O$_b$~2$p_y$ (O$_c$~2$p_x$) orbital of them hybridizes with the V~3$d_{xy}$ orbital in the $\bf a$ ($\bf b$) direction the anisotropy in the behavior of $\sigma$($\omega$) right after the absorption threshold has its origin in the different probabilities for an electron to hop from the O$_b$~2$p_y$ (O$_c$~2$p_x$) orbital into the V~3$d_{xy}$ orbital. The effective V~3$d_{xy}$--O$_b$~2$p_y$ (1.75~eV) and V~3$d_{xy}$--O$_c$~2$p_x$ (0.55~eV) hopping terms were obtained by adjusting the parameters of a tight binding (TB) model so as to reproduce the calculated dispersion of the lowest unoccupied bands originating from the V~3$d_{xy}$ states (for the detailed description of the TB model and the parameter values for $\alpha '$--NaV$_2$O$_5$ see Ref. \onlinecite{nav2o5tb}). The V3$d_{xy}$--O$_b$~2$p_y$ term is responsible for the splitting of the four bands into two sub-bands leading to the opening of the additional gap, mentioned above. As the V--O$_b$ distance in V$_2$O$_5$ is smaller than in $\alpha '$--NaV$_2$O$_5$ this hopping is slightly larger than the value of 1.7~eV found for the latter. The V~3$d_{xy}$--O$_c$~2$p_x$ term, which governs the dispersion of the bands along the $\Gamma-Y$ direction, is, however, smaller than the corresponding term in $\alpha '$--NaV$_2$O$_5$ (1~eV), despite the smaller V--O$_c$ distance in V$_2$O$_5$. It should be pointed out that in order to minimize the number of parameters the O$_c$~2$p_y$ orbitals were not included in the TB model. Their effect was taken into account implicitly via renormalization of other hopping terms. Bearing this in mind, the reduction of the effective V~3$d_{xy}$--O$_c$~2$p_x$ hopping, which reflects the weaker dispersion along the $\Gamma-Y$ direction, can be explained by an increased contribution of the V~3$d_{xy}$--O$_c$~2$p_y$ hopping of opposite sign caused by a larger displacement of O$_c$ atoms out of V--V chains running along the $\bf b$ direction. If the oxygen octahedra were undistorted the O$_c$~2$p_y$ orbitals would be orthogonal to the V~3$d_{xy}$ orbitals and would affect the dispersion only via the renormalization of the effective V~3$d_{xy}$--V~3$d_{xy}$ hopping. However, as the distortion increases the O$_c$~2$p_y$ orbitals start to overlap with the V~3$d_{xy}$ orbitals, thus suppressing the effective V~3$d_{xy}$--O$_c$~2$p_x$ hopping.

The hatched area in Fig. \ref{Sigmas}(a) shows the contribution from transitions into final states with energies 3.06~eV $\leq$ E$_f$ $\leq$ 4.4~eV and thus represents the transitions into the V~3$d_{xz}$/3$d_{yz}$ states having an admixture of mainly O$_v$~2$p_x$ and O$_v$~2$p_y$ character. 
In $\sigma_{xx}$ this contribution appears with more or less constant intensity in the energy region between 3~eV and 8~eV reflecting the flatness of the total density of occupied states (not shown here). However, in $\sigma_{yy}$ these transitions dominate within two separated energy intervals at about 4.8 and 8~eV. In the $\bf c$ direction they almost fully determine the absorption in the region between 3 and 7~eV. The part of the spectra which is due to transitions to vanadium $e_{g}$ states mixed with O$_b$~2$p_x$, O$_c$~2$p_x$/2$p_y$, O$_v$~2$p_z$ derived bands (4.4~eV $\leq$ E$_f$ $\leq$ 6.8~eV) is shown as the area patterned with vertical lines. It is clear that these transitions are responsible for the high intensity regions in $\sigma_{xx}$ and $\sigma_{yy}$ at about 6.6~eV and they dominate $\sigma_{zz}$ in the energy region between 7 and 12~eV due to the strong overlap between the O$_v$~2$p_z$ and V~3$d_{3z^2- r^2}$ orbitals. 

We complement the discussion about the strong optical anisotropy by presenting calculations of the optical conductivity with the O~3$d$ orbitals excluded from the LMTO basis set. 

The results (see dotted grey lines in Fig. \ref{Sigmas}(b)) show a strong decrease of the calculated $\sigma$ in the energy range up to 10~eV. This illustrates the important role of the O~$2p \rightarrow $~O~3$d$ transitions in the matrix elements. The calculations were repeated with O~$3d$ states of only one type of inequivalent oxygen atoms included in the basis. From the comparison of these spectra with those calculated with the maximal basis set one can see that $\sigma_{zz}$ and $\sigma_{yy}$ are effected mainly by O~$2p \rightarrow$~O~3$d$ transitions arising from O$_v$ and O$_c$ atomic spheres, respectively, which is illustrated by the black solid lines in the upper and middle panel of Fig. \ref{Sigmas}(b). However, the increase of the intensity of $\sigma_{xx}$ is determined to a great extent by O$_b$~$2p \rightarrow $~O$_b$~3$d$ transitions with an important contribution from O$_c$ near the absorption threshold and above 5.5~eV (see the black solid and the black dashed line in the lower panel of Fig. \ref{Sigmas}(b)). 

The analysis of the orbital projected O~3$d$ DOS (not shown) shows that only O$_b$ 3$d_{xy}$, O$_c$ 3$d_{x^2-y^2}$ and, to a lesser extent, O$_c$ 3$d_{xy}$ orbitals contribute to the wave functions of the lowest unoccupied bands formed by V~3$d_{xy}$ states. Thus, due to the dipole selection rules the $z$ components of the optical matrix elements for all transitions
from the O~2$p$ levels into these O~3$d$ states are zero which explains the negligible contribution to $\sigma_{zz}$ from the interband transitions into the final states in the energy range up to 3.06~eV. The high intensity of $\sigma_{xx}$ just above the threshold arises from a superposition of O$_b$~2$p_y \rightarrow$~O$_b$~3$d_{xy}$ and O$_c$~2$p_x \rightarrow$~O$_c$~3$d_{x^2-y^2}$ contributions due to the large spectral weight of O$_c$~2$p_x$ and especially of O$_b$~2$p_y$ states near the top of the valence band. Finally, the shift of the spectral weight of the low energy peak in $\sigma_{yy}$ to higher energies, compared with its counterpart in $\sigma_{xx}$, can be explained by the fact that the DOS of O$_c$~2$p_y$ and O$_b$~2$p_x$ states reach their maxima $\sim$ 1~eV below the top of the valence band and rapidly decrease at higher energies (see Fig. \ref{DOS}(b)). 

It should be noted that the optical matrix element depends not only on the contributions from individual spheres but also on the expansion coefficients which determine the weight of the orbital in the Bloch wave functions of the initial and  the final states and represent the symmetry of the crystal. Taking into account the exact form of the wave functions  can result in a suppression of some contributions allowed by the selection rules inside a sphere but makes the analysis very complicated and is thus beyond the scope of this paper.

\section{Summary}
\label{concl}

In conclusion, we have carried out $\bf q$-dependent, high resolution EELS measurements of V$_2$O$_5$ single crystals. The measured dynamical response showed no distinct dispersion along the principal crystallographic axes indicating the local character of the final states. The results of our LMTO calculations of the electronic structure agree well with previous LDA based numerical studies and can be easily interpreted in terms of bonding-antibonding and ligand field splitting, considering the crystal structure of V$_2$O$_5$ as a 3D network of corner and edge-sharing VO$_6$-octahedra. The optical conductivity, calculated from the energy bands and the LMTO eigenvectors show an excellent agreement with that derived from the experimental loss spectra in a wide energy range. The divisibility of the DOS of the final states into energy regions with contributions from states with certain symmetries and the calculation of the optical conductivity excluding the O~2$p \rightarrow $~O~3$d$ transitions for the inequivalent oxygen positions allowed us to explain the emergence of the main peaks as well as the observed anisotropy in the optical spectra. Because of the structural resemblance we expect a similar dynamical behavior of the O~2$p$ electrons in the quarter-filled compound $\alpha '$--NaV$_2$O$_5$. However, the higher probability for the effective V~3$d_{xy}$--O$_c$~2$p_x$ hopping in the latter compound points to an enlarged mobility of the O~2$p$ electrons along the legs.  Nevertheless, the knowledge gained about the dynamics of the O~2$p$ electrons of vanadium pentoxide provides us with a good starting point for the investigation of the optical properties of $\alpha '$--NaV$_2$O$_5$.

\acknowledgments
SVB gratefully acknowledges the hospitality extended to him during his stay at the Institute for Solid State and Materials Research in Dresden. This work was supported by the SMWK (4-7531.50-040-823-99/6) and the DFG (Fi439/7-1).

\begin{figure}
\caption{Crystal structure of V$_2$O$_5$. Vanadium atoms are shown as black spheres (upper panel). Bridge oxygens - dark grey; vanadyl oxygens - light grey; chain oxygens - white. 3D-network of distorted VO$_6$ octahedra (lower left panel) . 2D-network of VO$_5$ pyramids (lower right panel).}
\label{Structure}
\end{figure}

\begin{figure}
\caption{Electron energy-loss spectra measured with momentum transfer $\bf q$ parallel to (a) the crystallographic $\bf a$ and (b) the crystallographic $\bf b$ direction.}
\label{Lossfunc}
\end{figure}

\begin{figure}
\caption{ (a) Partial vanadium densities of states (DOS) of V$_2$O$_5$. (b) Partial oxygen DOS of V$_2$O$_5$. The dashed lines in the upper panels show the contributions of the O~3$d$ states (multiplied by 10) to the total DOS. Partial DOS are given per atom. Energies are given relative to the valence band maximum E$_V$. }
\label{DOS}
\end{figure}

\begin{figure}
\caption{The optical conductivity $\sigma_{xx}$ and $\sigma_{yy}$ of V$_2$O$_5$ determined by LDA band-structure calculations compared with those derived from the experiment. Theoretical curves are broadened with a Lorentzian of 0.3~eV width and shifted by 0.35~eV to higher energies.}
\label{Th+exp}
\end{figure}

\begin{figure}
\caption{(a) Decomposition of the calculated optical conductivity into three parts which represent the interband transitions into selected energy intervals (E$_f$) of the conduction bands: grey area: total optical conductivity, area with horizontal lines: 0~$\leq$ E$_f <$~3.1~eV, hatched area: 3.1~$\leq$~E$_f <$~4.4~eV, area with vertical lines: 4.4~$\leq$~E$_f <$~6.8~eV. (b) Optical conductivity calculated as in (a) but excluding successively the O~2$p\rightarrow$~O~3$d$ transitions at the three inequivalent oxygen positions: 
the solid grey line and the dotted grey line in each panel represent the optical conductivity, including all and none of the transitions into the O~3$d$ states, respectively. While $\sigma_{zz}$ and $\sigma_{yy}$ are dominated by O~2$p\rightarrow$~O~3$d$ transitions on O$_v$ and O$_c$, respectively (black solid lines in the upper and the middle panel), the contributions to $\sigma_{xx}$ from transitions into the 3$d$ states of O$_c$ (black solid line in the lower panel) and O$_b$ (black dotted line in the lower panel) are almost equal.}
\label{Sigmas}
\end{figure}

\end{document}